\documentclass[prl,twocolumn,superscriptaddress,groupedaddress,nofootinbib]{revtex4}
\usepackage{color}
\usepackage{graphicx}
\usepackage{dcolumn}
\usepackage{bm}
\usepackage{amssymb}
\usepackage{amsmath}
\usepackage{colortbl}
\usepackage{latexsym}
\usepackage{float}
\usepackage{ifthen}
\usepackage{caption,subfig}
\usepackage{enumerate}
\usepackage{url}
\usepackage{caption,subfig}

\def\ba{\begin{eqnarray}}
\def\ea{\end{eqnarray}}
\def\mpl{M_{\rm Pl}}

\def\({\left(}
\def\){\right)}
\def\ie{{\it i.e. }}

\def\nn{\nonumber}
\def\p{\partial}
\def\mn{_{\mu \nu}}

\def\stu{St\"uckelberg }
\def\p{\partial}

\def\<{\langle}
\def\>{\rangle}

\def\d{\mathrm{d}}

\def\gd{g_{\rm eff}}
\def\gu{g^{\rm eff}}
\def\geff{\gamma_{\rm eff}}

\newcommand{\para}[1]{\par\vspace{2mm}\noindent{\bf\emph{{#1}}}.---}

\begin{document}

\title{Ghosts \& Matter Couplings in Massive (bi-\&multi-)Gravity}
\author{Claudia de Rham}
\address{CERCA \& Department of Physics, Case Western Reserve University, 10900 Euclid Ave, Cleveland, OH 44106, USA}
\author{Lavinia Heisenberg}
\address{Nordita, KTH Royal Institute of Technology and Stockholm University,
     Roslagstullsbacken 23,  10691 Stockholm, Sweden}
\address{Department of Physics \& The Oskar Klein Centre,
AlbaNova University Centre, 10691 Stockholm, Sweden
}
\author{Raquel H.~Ribeiro}
\address{CERCA \& Department of Physics, Case Western Reserve University, 10900 Euclid Ave, Cleveland, OH 44106, USA}
\date{\today}

\begin{abstract}
 Recently, several works have investigated the coupling to matter in ghost-free massive \mbox{(bi- \& multi-)}gravity and a new effective coupling to matter has been proposed.
 In this note we clarify some confusion on the existence and the implications of a ghost above the strong coupling scale. We confirm that the standard constraint which is otherwise typically present in this type of theories disappears on generic backgrounds as soon as this new coupling is considered. This implies the re-emergence of the Boulware--Deser ghost. Nevertheless the absence of ghost in the decoupling limit implies that the cut-off scale (if identified with the scale at which the ghost enters) is higher than the strong coupling scale. Therefore there is a valid interesting region of applicability for these couplings at scales below the cut-off.
\end{abstract}

\maketitle

\para{Introduction}
Massive gravity \cite{deRham:2010ik,deRham:2010kj}, bi-gravity \cite{Hassan:2011zd} and multi-gravity \cite{Hinterbichler:2012cn} (see Ref.~\cite{deRham:2014zqa} for a recent review) share in common the fact that their very formulation involves several metrics. While it is natural to couple these theories to matter in the same way as in General Relativity (GR), there are a number of possible couplings to matter which have been explored in the literature, and can lead to interesting new phenomenology \cite{Khosravi:2011zi,Akrami:2012vf,Akrami:2013ffa,Akrami:2014lja,Tamanini:2013xia}.

In all generality, considering $N$--interacting metrics $g^{(I)}\mn$, and working in the Einstein frame (where the kinetic term for each of these metrics is the Einstein--Hilbert one), one can consider the following possibilities:

{\bf 1.} Each metric can couple to its own separate matter sector,
\ba
\mathcal{L}_{\rm matter}=\sum_{I=1}^N \mathcal{L}_I\(g^{(I)}\mn,\p \psi^{(I)}, \psi^{(I)}\)\,,
\ea
where the $\psi^{(I)}$ symbolize all the matter fields in a given sector\footnote{In all generality, the matter Lagrangian could also include Galileon-type of interactions \cite{Nicolis:2008in} for the matter field which would involve $\p^2 \psi^{(I)}$.}. In Ref.~\cite{Hassan:2011zd} such couplings to matter were shown to be free of the Boulware--Deser (BD) ghost \cite{Boulware:1973my} at all scales. The cosmology with this type of coupling to matter was studied for instance in \cite{Khosravi:2011zi,Akrami:2012vf,Akrami:2013ffa,Comelli:2011zm,Comelli:2012db,Comelli:2014bqa,DeFelice:2014nja,Konnig:2014dna,Aoki:2014cla} (a complete list of references is beyond the scope of this note and we refer the reader to \cite{deRham:2014zqa} for a review of many more efforts in this direction).

{\bf 2.} The same matter sector (symbolized by $\chi$) can couple to two or more separate metrics simultaneously as follows
\ba
\label{eq:TwoCouplings}
\mathcal{L}_{\rm matter}\supset \mathcal{L}_I\(g^{(I)}\mn, \p \chi, \chi \) + \mathcal{L}_J\(g^{(J)}\mn, \p \chi,  \chi \) \,,
\ea
with $I\ne J$ and where the kinetic term of the fields appears in both Lagrangians. The phenomenology of such couplings was considered in  \cite{Khosravi:2011zi,Akrami:2012vf,Akrami:2013ffa,Akrami:2014lja}. However such a coupling leads to a BD ghost  at an unacceptably low scale \cite{Yamashita:2014fga,deRham:2014naa}. While the arguments in \cite{Yamashita:2014fga,deRham:2014naa}  were made for massive (bi-)gravity, the appearance of the ghost for multi--gravity is implicit from these analyses.

  {\bf 3.}  An alternative to the previous coupling  \eqref{eq:TwoCouplings} where the same matter sector can couple to two or multiple metrics simultaneously was proposed in \cite{Yamashita:2014fga}, where the kinetic term of the field only couples to one metric while the potential may couple to several metrics, for instance
\ba
\label{eq:TwoCouplingsOther}
\mathcal{L}_{\rm matter}\supset \mathcal{L}_I\(g^{(I)}\mn, \p \chi, \chi \) + \mathcal{L}_J\(g^{(J)}\mn,  \chi \) \,.
\ea
Such couplings were shown to be free of the BD ghost classically in \cite{Yamashita:2014fga} but the ghost always reappears at an unacceptable  low scale at the quantum level \cite{deRham:2014naa}.

  {\bf 4.} Another alternative driven by the deconstruction framework was proposed in \cite{deRham:2013awa}, where different matter field sectors could interact through their potential, for instance as follows
\ba
\label{eq:DeconstructionCoupling}
\mathcal{L}_{\rm matter}=\sum_{I=1}^N \Big[\mathcal{L}_I\(g^{(I)}\mn,\p \psi^{(I)}, \psi^{(I)}\)\nn \\
-\mu_{I}^2 \sqrt{-g^{(I)}} \psi^{(I)}\psi^{(I+1)}\Big] \,.
\ea
However applying the same arguments as in \cite{deRham:2014naa} one can easily see that such couplings would also introduce a BD ghost at an unacceptable low scale at the quantum level unless the scales $\mu_I$ are well-below the lightest graviton mass scale, and thus unimportant.

{\bf 5.}  Finally the coupling to matter via an effective `composite' metric $\gd$ was considered in  \cite{deRham:2014naa},
\ba
\label{eq:effMetric}
\gu\mn=\alpha^2 g\mn +2 \alpha \beta g_{\mu \alpha}\(\sqrt{g^{-1}f}\)^\alpha_{\ \nu}+\beta^2 f\mn\,,
\ea
where $g\mn$ and $f\mn$ are two metrics (in bi-gravity they are the two dynamical metrics, while in massive gravity $g\mn$ can represent the dynamical metric and $f\mn$ can be the reference metric, for which Minkowski would then be a natural Lorentz-invariant choice). The two parameters $\alpha$ and $\beta$ are dimensionless and at this stage it is more convenient to keep them arbitrary so as to be able to dial either one of them to zero.

In terms of the effective metric \eqref{eq:effMetric}, the coupling to matter then takes the standard form,
\ba
\mathcal{L}^{\rm  massive (bi-)gravity}_{\rm matter}&=& \mathcal{L}_g\(g, \p \psi_g, \psi_g\)+\mathcal{L}_f\(f, \p \psi_f, \psi_f\)\nn \\
&+&\mathcal{L}_{\rm eff}\(\gd, \p \chi, \chi\)
\,,
\ea
where in the case of massive (bi-)gravity there can be three matter sectors, the first one, (symbolized by the fields $\psi_g$) only couples to the metric $g\mn$. The second sector (symbolized by the fields $\psi_f$) only couples to the metric $f\mn$, while the third sector (fields $\chi$) couples to both metric directly via the effective composite metric $\gd$ \eqref{eq:effMetric}. Of course quantum corrections will force every sector to couple to every metric, but such couplings will be highly suppressed and are thus unimportant for the rest of this discussion.

\para{Effective metric} In what follows we will focus on the last possibility, where a ghost is not generated from the integration of the matter fields alone. Instead integrating loops of matter fields simply generates one of the massive gravity `allowed' potentials derived in \cite{deRham:2010ik,deRham:2010kj}. This means however that the graviton mass ceases to be technically natural in this model unless only massless fields couple to $\gd$ (or unless $\gd$ couples to a supersymmetric sector or a sector endowed with a special symmetry or structure) \cite{deRham:2012ew,deRham:2013qqa}.

The generalization of the coupling to the effective metric \eqref{eq:effMetric} was also considered for multi-gravity in  \cite{Noller:2014sta}. In the vielbein formalism the effective vielbein takes the remarkable simple form
\ba
e_{\rm eff} = \sum_{I=1}^N \alpha_{I} e^{(I)}\,,
\ea
where $e^{(I)}$ is the vielbein for the metric $g\mn^{(I)}$, and the $\alpha_{I}$ are arbitrary dimensionless coefficients.

The coupling of matter to the effective metric \eqref{eq:effMetric} was considered  in the mini-superspace approximation in  \cite{deRham:2014naa} where it was shown to be free of the BD ghost in that approximation. This new coupling also allows for exact FLRW solutions which would otherwise be absent in massive gravity\footnote{We emphasize however that the absence of exact FLRW solutions for massive gravity does not prevent the existence of solutions which are arbitrarily close to FLRW within our horizon. On distance scales beyond our horizon, the Universe may well look inhomogeneous or anisotropic.} \cite{D'Amico:2011jj}. The absence of BD ghost was also proven around these exact FLRW solutions. We therefore emphasize that the existence of exact FLRW solutions cannot be due to ghostly operators since there are no such operators in the FLRW case.
Finally a decoupling limit analysis showed the absence of ghost at the strong coupling scale $\Lambda=(\mpl m^2)^{1/3}$ where $\mpl$ is the Planck scale and $m$ the graviton mass. That very fact alone shows that one can consider massive (bi-)gravity with that coupling to matter  as an EFT (effective field theory) at the very least till the strong coupling scale $\Lambda$, this is already sufficient to deduce certain phenomenological aspects from these couplings as was pointed out in \cite{deRham:2014naa}. A more thorough discussion on the implications of a ghost above the strong coupling scale will follow at the end of the paper.

Very recently, a more complete set of cosmological solutions were derived in Ref.~\cite{Enander:2014xga}. In particular J.~Enander {\it et.al.} showed that for certain choices of  parameters the background evolution is very similar (or even identical) to $\Lambda$CDM at early and late times. For other choices of parameters the equation of state parameter could differ from $-1$ and could have interesting signatures while still being in agreement with observations. Perturbations about these solutions are yet to be performed to determine their stability and their phenomenological viability, however the analysis performed in \cite{deRham:2014naa} already shows the absence of BD ghost about these solutions.

\para{Ghost-freedom?}
Recently it was claimed in \cite{Hassan:2014gta} that for a  coupling to matter of the form $\mathcal{L}_{\rm eff}\(\gd, \p \chi, \chi\)$, the Hamiltonian is linear  in the lapse (after integrating out the shift), and so the standard primary second-class  constraint which is present in massive (bi-)gravity is also present when this coupling is included and projects out the BD ghost at all energy scales below the Planck scale.

If correct this result would be of great interest. It would imply the total absence of BD ghost ({\it i.e.} all the way till the Planck scale) and extend the region of interest of this new coupling to any scale well beyond the strong coupling scale (so long as one does not treat the strong coupling scale as the cut-off -- see \cite{deRham:2014wfa} for a related discussion on that point).

While remarkable, this result raises however several physical questions. First the perturbative analysis performed in \cite{deRham:2014naa} showed the coupling to the effective metric introduces some non-linearities in the lapse  in the Hamiltonian (after integrating out the shift) which contradicts the claims of  \cite{Hassan:2014gta}. Second the new couplings to  matter can be written in Jordan frame at the price of changing the kinetic structure of the metric(s). This should then be put in light of the results presented in \cite{deRham:2013tfa}, for which no new kinetic terms were found in massive gravity without a BD ghost at some scale.  (In that analysis it was implicitly assumed that matter coupled in the standard way and the \stu fields were introduced accordingly).

In what follows we present a loophole behind the arguments of  \cite{Hassan:2014gta} which unfortunately invalidates the proof  for the total absence of BD ghost at all scales. We also present the details behind the analysis presented in  \cite{deRham:2014naa} which are sufficient to prove the existence of a ghost at a scale between the strong coupling scale and the Planck scale. We end this note with a discussion on the relevance of ghosts above the cut-off in EFTs.

\para{$(1+1)$--massive gravity}
The analysis performed in \cite{Hassan:2014gta} applies to any number of dimensions and as much for bi-gravity than for massive gravity. In order to focus on the essential points of the analysis and not get distracted with unnecessary formalism, we shall focus in what follows to the $(1+1)$-dimensional case of massive gravity with flat Minkowski reference metric where all the features of the arguments are already present. Any bi-gravity theory in any number of dimensions (greater or equal to two) can reduce to that simple $(1+1)$-massive gravity case in some limit.  This means that if this $(1+1)$ case diagnoses the existence of operators that are non-linear in the lapse, such operators must also be present in four-dimensional massive (bi-)gravity. If a BD ghost is present in the $(1+1)$-massive gravity case, then a BD ghost is also present in four-dimensional massive \mbox{(bi-)}gravity.

\para{ADM decomposition}
With these considerations in mind, we thus focus on the $(1+1)$-massive gravity case where the very same formalism as that derived in \cite{deRham:2010kj} applies. Writing the dynamical metric in the ADM form \cite{Arnowitt:1962hi} as
\ba
\d s^2_g=g\mn\d x^\mu \d x^\nu = - N^2 \d t^2+\(\d x+ N^1 \d t\)^2\,,
\ea
where for simplicity we have set the spatial metric to $\gamma_{11}=1$ (this is merely a matter of simplifying the derivation as much as possible and focusing on the essential points).
Following the same trick as that introduced in \cite{deRham:2010kj} and redefining the shift as follows
\ba
\label{eq:NewShift}
N^1=(1+N)n^1\,,
\ea
the contributions to the Hamiltonian from both the kinetic and mass term of the graviton are then linear in $N$ and take the very specific form \cite{deRham:2010kj},
\ba
\label{eq:NormalH}
\mathcal{H}_{\rm GR}+\mathcal{H}_{\rm mass} = \mathcal{A} N +\mathcal{B}
+(1+N)f(n^1)\,,
\ea
where $\mathcal{A}$ and $\mathcal{B}$ are independent of the lapse and the shift, and $f(n^1)$ is a function of the redefined shift only. (In the general case, $\mathcal{A}$ and $\mathcal{B}$ are functions of the spatial metric uniquely and $f(n^1)$ is a function of the spatial metric and the redefined shifts). If these were the only contributions to the Hamiltonian, we could immediately infer that the Hamiltonian remains linear in the lapse $N$ after integration of the shift and the lapse therefore  propagates a primary second class constraint which is sufficient to remove (at least half of) the BD ghost (since the theory is parity-invariant, a secondary constraint must then project out the rest of the ghost).
This is the essence of the argument behind the absence of ghost in massive (bi-)gravity.

\para{Matter Hamiltonian}
We now include the contributions from the matter field coupled to the effective metric \eqref{eq:effMetric}. For sake of simplicity it is sufficient to consider a massless scalar field $\chi$
\ba
\mathcal{L}_{\rm matter}=-\frac 12 \sqrt{{\rm det}\gd}\  \gd^{\mu\nu}\p_\mu \chi \p_\nu \chi\,.
\ea
This matter Lagrangian is of course not linear in the lapse (and neither should it be), however its contribution to the Hamiltonian is
\ba
\label{eq:Hmatter}
\mathcal{H}_{\rm matter}=\frac 12 \frac{N_{\rm eff}}{\sqrt{\gamma_{\rm eff}}}\(p_\chi^2 + (\p_1 \chi)^2 \)+ N^1_{\rm eff} p_\chi \p_1 \chi\,,
\ea
where $p_\chi=\p_{\dot \chi} \mathcal{L}_{\rm matter}$ is the conjugate momentum associated with $\chi$ and $N_{\rm eff}$, $N^1_{\rm eff}$, $\gamma_{\rm eff}$ are the respective effective lapse, shift and spatial metric of $\gd$
\ba
\label{eq:gamma_eff}
\gamma_{\rm eff}&=& \alpha^2+2\alpha \beta \xi  +\beta^2\\
\sqrt{\gamma_{\rm eff}}N_{\rm eff}&=& \alpha \beta (1+N) \xi^{-1}+(\alpha^2 N+\beta^2) \\
\gamma_{\rm eff}\, N^1_{\rm eff}&=& \alpha\, n^1 (\alpha (1+N)+2 \beta \xi) \,,
\ea
with $\xi=\(1-n_1^2\)^{-1/2}$. All these quantities are linear in the lapse $N$, in complete agreement with \cite{Hassan:2014gta}. This means that \eqref{eq:Hmatter} and thus the whole Hamiltonian $\mathcal{H}=\mathcal{H}_{\rm GR}+\mathcal{H}_{\rm mass}+\mathcal{H}_{\rm matter}$ is also linear in the lapse, and so far this is also in complete agreement with \cite{Hassan:2014gta} (of course when applied to the Hamiltonian as opposed to the Lagrangian or action).
Notice however that unlike in \eqref{eq:NormalH}, the new contribution  \eqref{eq:Hmatter} to the Hamiltonian depends on the shift in a way which is not factorizable, rather the dependence in the shift is of the following form
\ba
\mathcal{H}_{\rm matter} = g_1(n^1) (1+N) + g_2 (n^1)\,,
\ea
with $\p_n^1 \(g_1(n^1)/ g_2(n^1)\)\ne 0$. This means that the equation of motion for the shift {\it now does involve} the lapse
\ba
\frac{\delta \mathcal{H}}{\delta n^1}= (1+N) \( f'(n^1) + g_1'(n^1)\) + g_2'(n^1)=0\,.
\ea
Since the quantities $g_{1,2}$ involve the scalar field $\chi$ while $f$ does not, setting $g_2'(n^1)=0$ does not imply $ \( f'(n^1) + g_1'(n_1)\)=0$ and vise-versa. Instead the solution for $n^1$ does involve the lapse in a very non-trivial way and so when plugged back into the Hamiltonian, the latter will no longer be linear in the lapse. Since we will integrate out the shift, the exact field definition we choose for it (be it $N^1$, $n^1$ or anything else related locally to $N^1$) is irrelevant. The change of variable \eqref{eq:NewShift} is merely a matter of convenience and there is no notion of correct or incorrect field to work with.

\para{Shift equation}
In \cite{Hassan:2014gta} it was also found that the Hamiltonian was linear in the lapse prior integrating over the shift. From that fact it was deduce that the Hamiltonian ought to remain linear in the lapse also after integration over the shift. The argument went as follows (see Eq.~(4.15) of Ref.~\cite{Hassan:2014gta} and arguments thereafter):
The equation of motion with respect to the shift can be written as follows:
\ba
0&=&\frac{\delta \mathcal{H}}{\delta n^1}= (1+N) \( f'(n^1) + g_1'(n^1)\) + g_2'(n^1)\nn\\
&=& \frac{\delta \mathcal{H}}{\delta N^1}\frac{\delta N^1}{\delta n^1}= \frac{\delta \mathcal{H}}{\delta N^1}  (1+N)\,.
\ea
Since the RHS term on first line is linear in the lapse and since the Jacobian factor $\delta N^1/\delta n^1 = (1+N)$ is also linear in the lapse,  Ref.~\cite{Hassan:2014gta}  concluded that the contribution $\delta \mathcal{H}/\delta N^1 $ (when expressed in terms of $n^1$) also ought to be independent of the lapse ``otherwise there would be nonlinear terms" (in the lapse on the RHS of the second line).

If this argument was correct, it would imply the absence of ghost for any theory whose Hamiltonian is linear in the lapse after a redefinition of the shift which is also linear in the lapse. Consider for instance the redefinition of the shift introduced in \eqref{eq:NewShift} and let $\mathcal{L}_1$ be a Lagrangian whose Hamiltonian $\mathcal{H}_1$ is linear in the lapse when expressed in terms of $n^1$. Then according to the previous argument of  Ref.~\cite{Hassan:2014gta}, this Lagrangian $\mathcal{L}_1$ is free of the BD ghost at all scales. Now consider a second Lagrangian $\mathcal{L}_2=\mathcal{L}_1- m^2 \mpl^2 (N^1)^2/(1+N)^2$. The associated Hamiltonian for this second theory is then $\mathcal{H}_2=\mathcal{H}_1+m^2\mpl^2 (n^1)^2$ and is also linear in the lapse (when expressed in terms of the redefined shift $n^1$). So according to the previous argument we would infer that both theories $\mathcal{L}_1$ and $\mathcal{L}_2$ are free of the BD ghost. However this cannot be so. The new operator $\mathcal{L}_{\rm new}=-m^2\mpl^2 (N^1)^2/(1+N)^2$ in $\mathcal{L}_2$ can be expressed in terms of the \stu fields and includes a contribution of the form $\mathcal{L}_{\rm new} \supset \frac{1}{\mpl m^4} (\p_1 \dot \pi)^2 \ddot \pi$, where $\pi$ is the helicity-0 mode of the graviton. Such an interaction would carry a ghost at the scale $\Lambda_5= (\mpl m^4)^{1/5}\ll \Lambda$ unless it was canceled by other terms in $\mathcal{L}_1$. In conclusion not both $\mathcal{L}_1$ and $\mathcal{L}_2$ can be ghost-free.

The resolution behind this apparent discrepancy lies in the fact
$\delta \mathcal{H}/\delta N^1$ (when expressed in terms of $n^1$) can actually  depend on the lapse in the following way $(k_1(n^1)N+k_2(n^1))/(1+N)$. An explicit computation of $\delta \mathcal{H}/\delta N^1$ shows that this is indeed what happens in the case of the coupling to the effective metric and $k_1\ne k_2$ as soon as $\alpha\beta \ne 0$. This implies that the equation of motion for the shift does depend non-trivially on the lapse as found in \cite{deRham:2014naa}.

\para{Integrating out the shift}
We now proceed with deriving the equation for the shift. For that we consider the following Hamiltonian (see \cite{deRham:2010kj} where for simplicity we may ignore the contributions in the mass term which do not involve the shift)
\ba
\mathcal{H}=N R^0 + (1+N) \(n^1 R^1+2 m^2 \sqrt{1-n_1^2}\)+\mathcal{H}_{\rm matter}\,,
\ea
where $R^0$ and $R^1$ do not depend on the lapse nor shift. Technically $R^0=R^1=0$ in this case but we keep them so as to show that our results is not an artefact of this two-dimensional case.
The equation of motion for the shift is then given by (writing $\chi'=\p_1 \chi$)
\begin{widetext}
\ba
2\geff^2\frac{\delta \mathcal{H}}{\delta n^1}&=& \Big\{
(2R_1-4 m^2 \xi n^1)\geff^2+2 \alpha^2 \chi' p_\chi (\geff-2\alpha \beta \xi^3 n_1^2)
-\alpha \beta \xi n_1 \(\geff+2\alpha \xi (\alpha \xi+\beta)\)\(p_\chi^2+\chi'{}^2\)
\Big\}(1+N)\nn \\
&+&2 \alpha \beta \xi^2 \Big\{
\xi n^1 (\alpha^2-\beta^2)\(p_\chi^2+\chi'{}^2\)+2\(2\alpha \beta + (\alpha^2+\beta^2)\xi\)p_\chi \chi'
\Big\} \equiv 0\,, \label{eq:shift}
\ea
\end{widetext}
where we bear in mind that both $\gamma_{\rm eff}$ given in \eqref{eq:gamma_eff} and $\xi$ are functions of $n^1$. We see immediately that when $\alpha \beta=0$, the second line of the previous equation vanishes and the dependence on the lapse completely factorizes out which means that this equation can then be solved without involving the lapse.  As soon as $\alpha \beta \ne 0$ this remarkable feature disappears. We emphasize that so far the equation \eqref{eq:shift} is exact and no perturbative expansion has been performed.

At this level the most straightforward way to proceed would be simply to solve the shift equation of motion \eqref{eq:shift} for the lapse in terms of the shift, the spatial metric and the scalar field (since this is now possible if $\alpha \beta\ne 0$) and plug that expression back in the Hamiltonian. It is then clear that the shift would not propagate a constraint for the spatial metric. In this 2d case, we would then be left with one physical degree of freedom in the spatial metric (had we tracked $\gamma_{11}$) and one physical degree of freedom in the scalar field, which is one too many degrees of freedom. In four dimensions, the same counting would go through and we would obtain six physical degrees of freedom for the massive graviton which corresponds to the five standard ones and the BD ghost as the sixth one.

However in order to avoid any confusion, let us solve \eqref{eq:shift} as an equation for the shift itself.
Independently of whether or not $\alpha \beta =0$ that equation is highly non-linear in the shift and many different branches of solutions exist. However none of them are independent of the lapse. Moreover one should really consider the branch of solution which is continuous with the flat case limit, \ie\ the one for which $n^1=0$ when $R_1=0=p_\chi=\chi'$. Performing a perturbation analysis as was done in \cite{deRham:2014naa} is a perfectly safe and well-defined procedure that will successfully identify a ghost about backgrounds which are connected to Minkowski (\ie for which Minkowski would be a solution in the absence of any source). Certainly this is a desirable feature of the theory. Nevertheless, a certain level of confusion as to the validity of perturbation theory exists in the literature and to avoid any possible source of concern, we take here a different approach.

Instead we solve the equation for the shift as an expansion in the lapse
and work with the branch of solution which admits a smooth limit when $\beta=0$.   When plugged back into the Hamiltonian, the latter takes a relatively simple form,
\ba
\label{eq:Hexp}
\mathcal{H}=\mathcal{H}_0+\mathcal{H}_1 N +\mathcal{H}_2 N^2+ \mathcal{O}(N^3)\,,
\ea
with $\mathcal{H}_{0,1}\ne 0$ and to second order in the parameter $\beta$ while keeping $\alpha=1$, one finds\footnote{One could equivalently set instead $\beta=1$ and expand in the coefficient $\alpha$ without affecting the essence of the result.}
\ba
\label{eq:H2}
\mathcal{H}_2 &=&\frac{\beta^2}{32 m^8}\mathcal{X}\left[(p_\chi^2+\chi'{}^2)(p_\chi \chi'+R_1)+2p_\chi \chi' \mathcal{X}\right]^2\nn \\
&+&\mathcal{O}(\beta^3)\,,
\ea
with
\ba
\mathcal{X}= \sqrt{4m^4+(p_\chi \chi'+R_1)^2}\,.
\ea
 When expanding the Hamiltonian $\mathcal{H}_2 N^2$ to sixth order in perturbations, (\ie to sixth order in $\epsilon$ with $p_\chi, \chi',R_1\sim \mathcal{O}\(\epsilon\)$ and $N\sim 1+ \mathcal{O}\(\epsilon\)$) we find precisely the same operator as pointed out in \cite{deRham:2014naa}.

\para{Implications for the BD ghost}
We have shown that when coupling to matter with the effective metric \eqref{eq:effMetric}, the Hamiltonian becomes non-linear in the lapse after integrating out the shift.
Next we can integrate out the lapse in the Hamiltonian. In the standard case, the lapse propagates a constraint, however as soon as the coupling is present (here since we set $\alpha=1$ this means as soon as $\beta\ne 0$, but in general it is as soon as $\alpha\beta \ne 0$) the equation of motion for the lapse should instead be solved for the lapse itself and imposes no constraint for the spatial metric nor for the field $\chi$. As a result in $d$ spacetime dimensions, there will be $d(d-1)/2$ physical degrees of freedom propagating in the metric and one in the scalar field. In four dimensions, this would correspond to 7 degrees of freedom which is one too many. Therefore we can conclude that the new coupling to matter does indeed generate a BD ghost on generic backgrounds as was found in \cite{deRham:2014naa}. With this in mind the physically interesting question one has to address is the scale at which BD enters. It will be important to establish whether that scale is infinitely close to $\Lambda$ or whether it is well separated from the strong coupling $\Lambda$.

While the analysis presented here applied for a massless scalar field coupled to the effective metric, a more general analysis along the lines of \cite{Hassan:2014gta} can be performed for an arbitrary matter sector. The crucial point of the previous analysis is the coupling to $N_{\rm eff}$ and $N^i_{\rm eff}$. Any matter field, in any number of dimensions, that has a kinetic term will involve a contribution proportional to $N_{\rm eff}/\sqrt{\det \geff}$ in the Hamiltonian. While $N_{\rm eff}/\sqrt{\det \geff}$ is linear  in the lapse, its dependence on the shift is not factorizable, so the conclusions of the previous section are not an artefact of the choice of a massless scalar field. The same results will hold for generic matter fields coupled to the effective metric \eqref{eq:effMetric}.


As emphasized earlier, even though this analysis was performed for massive gravity in $(1+1)$-dimensions, it will apply for any multi-gravity theory in any number of dimensions since these theories will always admit a limit where the previous derivations are applicable.

\para{EFTs with a ghost above their cut-off}
If a theory has a ghost within its very regime of interest, that theory is ill-defined (and sick) and no physics can be deduced from it. This means that if a theory has a ghost at a scale $M$, that theory can only be used as an EFT with cut-off $\Lambda_c\le M$.
However there appear to exist misconceptions in the literature that ghosts above the cut-off of an EFT could also  have (disastrous) effects on the EFT at scales below that cut-off. This is, however, impossible since by the very definition of an EFT, at energy scales below the cut-off, operators that enter above the cut-off (including ghostly operators) are small  and thus unimportant.
This is well understood already in the context of QED (see for instance Eq.~(63) of Ref.~\cite{Burgess:2007pt}), in gravity  (see for instance Ref.~\cite{Burgess:2003jk} where operators with higher orders in derivatives are considered at a scale above the cutoff) or  for instance in the context of inflation. See for instance Ref.~\cite{Creminelli:2010qf} for  a nice discussion and for examples of EFTs which include operators of the form $(\p^2 \pi)^3/\Lambda_c^5$, where $\Lambda_c$ is the cut-off of that EFT.

In fact most EFTs generate by quantum corrections operators which are ghostly but lie at or above the cut-off. This is the case for QED, QCD, GR, many models of inflation, Galileons, and of course for massive (bi-)gravity. So all these theories do really include ghosts above their cut-off even if these may not be explicitly seen at the classical level. However, so long as we restrict ourselves to scales below the cut-off, operators that enter at or above the cut-off can certainly not render the low-energy EFT unstable. In what follows we consider a few (four-dimensional) EFT examples to illustrate the fact that ghosts above the cut-off of EFTs are harmless.

$\bullet$ {\it A simple scalar field EFT:} As a first example consider the following EFT with cut-off $\Lambda_c$
\ba
\label{eq:SF1}
\mathcal{L}=-\frac 12 (\p \phi)^2 + \frac{1}{\Lambda^2_c}\(\Box \phi\)^2\,.
\ea
This EFT has a ghost at the scale $\Lambda_c$, but as long as one works at energy scales below $\Lambda_c$, meaning as long as the derivatives and the classical value of the field are small $\p \ll \Lambda_c$, $\phi\ll \Lambda_c$, we will always have $(\Box \phi)^2/\Lambda_c^2 \ll (\p \phi)^2$  and so that ghostly operator can be treated perturbatively compared to the normal kinetic term $(\p \phi)^2$ and one would not excite the associated Ostrogradsky ghost. See \cite{Weinberg:2008hq} for a review.

$\bullet$ {\it Multiple scales:} We now consider a case where mutiple scales are involved.  Consider the following EFT,
\ba
\label{eq:SF2}
\mathcal{L}=-\frac 12 (\p \phi)^2 + \frac{\beta}{\Lambda^4}\frac{(\p \phi)^4}{1+\Box \phi/\mpl^3}\,,
\ea
with the scale $\Lambda \ll \mpl$ and where the parameter $\beta$ is dimensionless (which could of course be set to one). The second term in $\mathcal{L}$ is a `package' that involves an infinite number of operators at different scales spanning from $\Lambda$ to $\mpl$. Some of these operators are ghostly but it certainly does not mean that this EFT only makes sense when $\beta=0$ or that physical predictions cannot be made and trusted with $\beta \ne 0$.

To be more precise this theory is strongly coupled at the scale $\Lambda$. It also has a Galileon-like operator at the higher scale $\tilde \Lambda= (\Lambda^4 \mpl^3)^{1/7}\gg \Lambda$. At the scale $M= (\Lambda^2 \mpl^3)^{1/5}\gg \tilde \Lambda$, this theory has a ghost and therefore the cut-off $\Lambda_c$ of this EFT should be $\Lambda_c \le M$.  We can work with this theory all the way up to the cut-off scale and excite not only the operators arising at the strong coupling scale $\Lambda$ but also  those arising at a higher scale $\tilde \Lambda$ and derive any relevant physical implications from this `package' of terms without ever needing to worry about the ghost.

When deriving classical solutions, or other physical implications from this  EFT,  one does not need to surgically remove the ghost terms from the package first. Working at energy scales below the cut-off will automatically ensure that the ghost is not excited since it cannot be excited at a scale below $M$, just like in the previous example.

$\bullet$ {\it Massive (bi-)gravity:} Massive (bi-)gravity is a (more complicated) version of the previous example. The decoupling limit analysis performed in \cite{deRham:2010ik} ensures that this theory has no ghost at its strong coupling scale $\Lambda$. This fact alone was sufficient to ensure that one could take the cut-off of that theory to be above $\Lambda$ and have an interesting regime of validity for the theory where the Vainshtein mechanism could be active. It was further proven that at the classical level that theory had no BD ghost all the way up to the Planck scale in \cite{Hassan:2011hr}, which means that the cut-off of the theory could actually be quite larger than $\Lambda$ (although it is still likely that the cut-off of the massive (bi-)gravity EFT will be below the Planck scale).

$\bullet$ {\it Massive (bi-)gravity with new couplings to matter:}  When the new couplings to matter are considered, we are again in a situation comparable to the second scalar field example.
This new coupling to matter can be seen as a package which includes an infinite number of operators entering at all scales between $\Lambda$ and $\mpl$, with a ghost above $\Lambda$.

The existence of a ghost in the theory is certainly a source of concern and one should ensure that this EFT is not used beyond its regime of validity. This means that this EFT can only be trusted at energy scales below the mass of the ghost. In \cite{deRham:2014naa} and \cite{Enander:2014xga,Schmidt-May:2014xla} exact solutions were found in the presence of this new coupling to matter. These can be trusted only at energy scales below the mass of the ghost as emphasized in \cite{deRham:2014naa}, but at these scales the solutions can indeed be trusted even if one has not surgically removed the pathological interactions that arise at and above the cut-off of this theory.  The exact value of the cut-off has not yet been fully determined and it should be established with care so as to evaluate whether or not this can have a full impact for cosmology.

The existence of exact FLRW solutions found in \cite{deRham:2014naa,Enander:2014xga} relies on the fact the scalar field $\chi$ equation of motion is now different and combined with the Raychaudhuri equation, no longer leads to a constraint for the scale factor.  In parallel, the decoupling limit analysis performed at the scale $\Lambda$  shows that the equation of motion for the scalar field $\chi$ is affected by this new coupling already at that scale (\ie differs from the equation one would infer if $\alpha\beta = 0$ in a way which involves the \stu fields). Since the ghost is not present at these scales, the modification in the scalar field equation cannot be caused (solely) by the BD ghost, and must include operators which do not involve the ghost. This realization plus the fact that both the FLRW background and its first order perturbations do not excite the BD ghost implies that the existence of exact FLRW in the presence of this new coupling to matter is not due  to the existence of a ghost but rather due to new operators which are healthy (and arise already at the scale $\Lambda$).  We emphasize once again that when imposing the FLRW symmetry, all the ghostly-like operators disappear, leading to a healthy theory on that background (and at first order in perturbations around it).

\para{When ghosts do matter} While ghosts above the cut-off are irrelevant for physics at energies below that cut-off,
it is worth stressing at this point that ghosts really do matter and have catastrophic effects when they enter within the regime of validity of the EFT. Fortunately there are many well-established methods to diagnose their existence, which we review here. In what follows we illustrate these methods with a few examples. The configurations~\eqref{eq:phiII} and \eqref{eq:H4} (with $\Lambda\gtrsim M$) represent examples of {\bf what we are not doing} when considering the coupling to matter with \eqref{eq:effMetric} in massive (bi-)gravity and exploring FLRW solutions at energies below the cut-off.\\

$\bullet$ First let us go back to our first scalar field example \eqref{eq:SF1}. As already mentioned, the EFT of \eqref{eq:SF1} is only valid at energy scales below $\Lambda_c$. When considering time-dependent classical solutions for \eqref{eq:SF1}, we may naively find two different branches of solutions. The first one being the trivial one $\phi_{\rm I}(t)\sim C_0 + C_1 t$, and the second one being a `self-sourced' or `self-accelerated' one
\ba
\label{eq:phiII}
\phi_{\rm II}(t)\sim e^{\pm \Lambda_c t}\,.
\ea
This second branch of solution relies on the existence of the ghostly operator $(\Box \phi)^2/\Lambda_c^2$ to exist and is not well-defined in the limit where that operator disappears. This second type of solution is not to be trusted. Fortunately one can directly see that this second class of solution necessarily involves derivative scales which are at or beyond the cut-off scale. Indeed $\p \sim \p \phi/\phi\sim \Lambda_c$ for this second type of solutions. So one can look at all the solutions of the Lagrangian \eqref{eq:SF1}, keeping the ghostly operator and we are still able to diagnose that the second type should be discarded straight away since it lies at an energy scale which is {\bf not below the cut-off}. This is why it is so important to make sure that the theory is not taken beyond its regime of validity. So long as we only consider solutions for which all the field and all its derivatives are smaller than the cut-off, we are ensured that the ghost plays no relevant role in these solutions.\\

$\bullet$ Next, consider the following gravitational theory,
\ba
\label{eq:ModGR}
{\mathcal{L}}=\frac{\mpl^2}{2}\sqrt{-g}\left[
R+\frac{\alpha}{12^2} R \(\frac 1 {\Lambda^4} R^2+\frac{4}{M^4}R\mn^2\)
\right]\,,
\ea
where $R$ is the standard scalar curvature and $R\mn$ the Ricci tensor. In this case the contribution from $R^3$ leads to extra degrees of freedom (namely a scalar field when this action is written in Einstein frame). The operator $R R\mn^2$ is ghostly. The existence of these extra degrees of freedom can already be established by performing an ADM analysis on FLRW (\ie in the mini-superspace approximation).
This means that one can derive classical solutions for this full theory (keeping the term $R R\mn^2$), but in order to trust these solutions one needs to ensure that all the curvature scales involved are smaller than $M$.

Focusing on exact FLRW solutions, we can find several branches of them. The first one is the trivial one $H^2=0$ in the absence of matter field ($H$ being the Hubble parameter). The Lagrangian \eqref{eq:ModGR} also admits a second type of `self-accelerated' solutions:
\ba
&&{\rm Branch\ I}:\quad  H=0\\
\label{eq:H4}
&&{\rm Branch\ II}:\quad  H^4= \frac{1}{\alpha}\frac{M^4 \Lambda^4}{M^4+\Lambda^4}\,.
\ea
Clearly this second type of solution does not exist in the limit where $\alpha\to0$. However it does not mean that this solution is pathological. As seen earlier, and emphasized throughout this manuscript and in \cite{deRham:2014naa}, as long as the scales involved are smaller than $M$, one can trust this solution.
\begin{enumerate}
\item[II$_a$]: In the case where $\Lambda\ll M$, one has $H \sim \Lambda \ll M$ and this solution is within the regime of validity of the EFT. One can check explicitly that we would have obtained the same solution as if we had removed the ghostly operator $R R\mn^2$ from the Lagrangian \eqref{eq:ModGR} prior to derive the classical solutions.
\item[II$_b$]: In the case where $\Lambda \gtrsim M$, one has $H \sim \Lambda \gtrsim M$, and clearly the classical solution \eqref{eq:H4} ought to be ignored in that case. However even without applying this criterion, one could consider the solution \eqref{eq:H4} and simply perform perturbations about it. Then a first order linear analysis will directly diagnose a ghost and other instabilities when $\Lambda \gtrsim M$ which are sufficient to signal that that solution is not to be trusted.
\end{enumerate}

In the case of the exact FLRW solutions found in \cite{deRham:2014naa} the situation is actually even better than the case II$_a$. First there are no ghostly operators in the mini-superspace approximation. This means that no FLRW solution can ever excite the ghost in the first place. Second, unlike for $f(R)$-gravity there are not even new degrees of freedom on FLRW when the new couplings to matter are considered. Finally let us stress that even if any doubts still existed on the validity of the exact FLRW solutions found in \cite{deRham:2014naa}, a perturbed analysis has been performed about the solutions and it was found that no ghosts were present, unlike what one would find in case II$_b$ of the previous example.

\para{Discussion}
When considering gravitational theories with many metrics, one is inclined to treat every metric on an equal footing. This is achieved at the level of bi-gravity when no coupling to matter is considered. As soon as these theories couple to matter it has to be done in a way which would break the equivalence between both metrics or a ghost will appear below the Planck scale (\ie the same matter sector cannot couple to both metrics in a symmetric way without introducing a BD ghost at a scale below the Planck scale).

The existence of such a ghost may or may not be disastrous depending on the setup. For instance if the same sector couples to two metrics as in \eqref{eq:TwoCouplings} then a ghost appears already at the strong coupling scale $\Lambda$ or even below and the regime of validity of the theory is rather limited \cite{Yamashita:2014fga,deRham:2014naa}. If the same sector couples to both metrics via an effective `composite' metric, as proposed in Ref.~\cite{deRham:2014naa} a ghost is also present in the spectrum but above the strong coupling scale $\Lambda$. The precise cut-off of this theory, or scale at which the ghost enters should be understood in more depth. However, this theory still provides useful physical predictions at energies below that cut-off, like any other EFT.\\

{\bf Acknowledgments:}
We would like to thank A.~Emir G\"umr\"uk\c c\"uo\u{g}lu, Andrew Matas, Johannes Noller and Andrew J.~Tolley for useful discussions.
CdR and RHR are supported by a Department of Energy grant DE-SC0009946.

\bibliography{references}

\end{document}